\documentclass[aps,prc,twocolumn,showkeys,showpacs,groupedaddress,superscriptaddress]{revtex4}
\usepackage{amssymb}
\usepackage{graphicx}
\usepackage{color}

\def\bra{\langle}
\def\ket{\rangle}
\def\eps{\epsilon}
\def\mev{\;{\rm MeV}}

\begin{document}

\title{Dumbbell shapes in the super-asymmetric fission of heavy nuclei}

\author{ F.A. Ivanyuk}
\email{ivanyuk@kinr.kiev.ua}
\affiliation{Institute for Nuclear Research, 03028 Kiev, Ukraine}

\author{N. Carjan}
\email{carjan@theory.nipne.ro}
\affiliation{Joint Institute for Nuclear Research, 141980 Dubna, Moscow Region,
Russia}
\affiliation{National Institute for Physics and Nuclear Engineering "Horia-Holubei", Reactorului 30, RO-077125, POB MG6,   Magurele-Bucharest, Romania}
\date{today}

\begin{abstract}
We have calculated the fission fragments' mass distributions for several isotopes of heavy and super-heavy nuclei from uranium to flerovium within an improved scission point model. For all considered nuclei, in addition to the standard mass-asymmetric fission mode we have found the mass super-asymmetric mode with the mass of heavy fragments
equal 190.
For the actinide nuclei, the probability of super-asymmetric fission is by 6 orders of magnitude smaller than for standard asymmetric fission. For the superheavy nuclei
this probability is only by 2 orders of magnitude smaller.
In all cases, the super-asymmetric scission shapes are dumbbells with the heavy
fragment close to a sphere. We have
estimated the stability of the light fragment concerning the variation of the neck and found out that sequential ternary fission is not favored energetically.
The calculations were carried out with nuclear shape described by generalized Cassinian ovals with 6 deformation parameters, $\alpha, \alpha_1, \alpha_2, \alpha_3, \alpha_4$ and $\alpha_5$. The configuration at the moment of the neck rupture was defined by fixing $\alpha=0.98$.  This value corresponds to a neck radius $r_{neck}\approx$ 1.5 fm.
\end{abstract}


\maketitle

\section{Introduction}
\label{intro}

The discovery of nuclear fission \cite{HS} and its almost simultaneous
qualitative
explanation \cite{BW} based on an already existing analogy between a nucleus
and a liquid drop (LD)\cite {Weizsaker} took place 85 years ago.
The first real advance in theoretical understanding of the fission process came
25 years later when it was shown that a nucleus, in addition to its
macroscopic LD energy has a microscopic energy called shell correction
\cite{Strutinsky}. It is this fluctuating energy that is responsible, among
others, for the division of a heavy nucleus
into two unequal fragments, as the experimental data have shown from the very
beginning. More exactly, this observed mass asymmetry is due to the extra stability
brought by the double-magic $^{132}$Sn to the heavy fragment group.
The explanation of this main asymmetric mass division in low
energy fission was at the origin of the revival of the research in the field of nuclear fission in the 70's \cite{VH}.

Theoretical and experimental investigations of the fragment mass distribution far
away from the main peak represented the exotic aspect of this new wave. Arguments
were brought in favor of an increase in the mass yield at very large asymmetries.
This type of mass division was called super-asymmetric fission (SAF).

For instance, calculations in the frame of a
simplified macroscopic-microscopic model \cite{Sandulescu1978,Lustig1980} based on
the two-center
shell model with only two shape parameters (elongation and mass asymmetry) predicted
a shoulder at A$_L$=60 with an yield 3$\times$10$^{-5}$ in $^{}$Fm and 10$^{-4}$ in
$^{252}$No.

Recently, the super-asymmetric fission events in super-heavy elements  were found in the Langevin calculations with the two-center shell-model shape parametrization \cite{our17}, Fourier-over-spheroid shape parametrization \cite{pasha}, in the
Brownian shape motion model \cite{BSM} and in the cluster radioactivity model \cite{poenaru2018,warda2018,cluster}.

The first experimental observation of an enhanced yields for $^{66}$Ni
and $^{67}$Cu was, by a radiochemical method, in fast-neutron fission of $^{238}$U \cite{Rao}. More recently \cite{knyaz} an increase of fragment yields in the
mass region around 52/208 was observed in the fission of $^{260}$No produced in
the reaction $^{12}$C +$^{242}$Cm.
At 50 MeV excitation energy, this yield was about 10$^{-3}$.

An important experimental campaign took place in the 90's at Lohengrin fission-fragment
mass separator (ILL-Grenoble). Very light fragments (from $A_L$=67 to 80)
were detected in five fissioning nuclei
(from $^{236}$U to $^{250}$Cf) using (n$_{th}$, f) reactions. This campaign ended
in 2004
\cite{Rochman} with the result: a shoulder in the mass yield (10$^{-4}$ to
10$^{-5}$) near
$A_L$=70 is observed in all studied cases. It was interpreted as SAF and attributed to
the stabilizing effect of $^{68}$Ni due to the subshell at N=40 \cite{Ni68,Ni68a}.
Since at Lohengrin one can detect only one fragment, it is not possible to know if
it originates from a binary or a ternary division.
One should mention that this shoulder is not present in  JEF2 library \cite{JEF2} or
in Wahl \cite{Wahl} evaluation.

An improved version of Wilkins scission point model (SPM) \cite{spm} was proposed recently \cite{CIOT} to calculate fission fragments mass and TKE distributions.
It uses pre-scission nuclear shapes described by generalized Cassini ovals.
Since the neck ruptures at finite radius ($\approx$ 2 fm) \cite{stlapo},
the distributions are  calculated just before scission and not when
the fragments are already separated as in the original model \cite{spm}.
Fission fragment observables were also calculated by a microscopic version of the scission-point model \cite{micro-spm}. The chosen scission configuration was taken from a
constrained Hartree-Fock-Bogoliubov model with Skyrme BSk27 interaction.

In the present work, we apply the above-mentioned improved SPM to the investigation of super mass-asym\-metric fission of heavy and super-heavy nuclei. It is simple and has remarkable achievements. For instance :

a) It has reproduced the bi-modal fission of heavy actinides. The double inversion of these two modes that occur with increasing mass A is, for the first time, clearly demonstrated to be the cause of the sharp transition from asymmetric to symmetric fission \cite{CIOT}.

b) It has obtained a surprisingly good agreement when comparing with existing data in all No and Rf isotopes \cite{cluster16}.

c) An incursion in the region of super-heavy elements (SHE) has revealed the crucial
role of the octupole deformation at scission in determining the main feature of the
mass distribution: symmetric or asymmetric \cite{CIO}. With $\alpha_3$, the division is
asymmetric and the position of the light fragment peak is constant at $A_L$=136.
In \cite{carjan2019} the excitation energy dependence of the fragment mass distributions in the fission of SHE was predicted.

We can therefore apply the model with confidence also to SAF.
The fission fragment mass distributions for several U, Pu, Fm, No,
 Hs and Fl isotopes are calculated
for a wide domain of fragment masses (from 40 to 210) to include
very large mass asymmetries.

In Section \ref{model} the main equation of the improved SPM is discussed.
 In Section \ref{potential} we present the mean-field model and the shape
 parametrization that we use.
 The calculation of the potential energy along the scission line is described in Section \ref{transp}.
In Section \ref{results} the results for the fission fragment mass distributions, the corresponding potentials and the average scission shapes of nuclei at the super mass-asymmetric peak are presented.
The probability that after scission the light fragment in SAF, to which the neck is attached, undergoes a 2nd fission is investigated in VI.
The average total kinetic energy of the fission fragments for all systems studied
here is estimated in \ref{kinetic}. TKE distributions for two nuclei that are susceptible to be measured, $^{256}$No are presented for standard asymmetric and super-asymmetric modes.
Section \ref{suma} contains the summary.
\section{The scission point model}
\label{model}
The main assumption in any SPM is that fission is a {\it slow} process and
that during the fission process, the state of the nucleus is close to
statistical  equilibrium, i.e.\
each point $q_i$ on the deformation energy surface is
populated with a probability given by the canonical distribution,
\begin{equation}\label{boltz} P(q_i)=e^{-\frac{E(q_i)-Z}{T_{coll}}},\,
Z\equiv -T_{coll}\log\sum_ie^{-\frac{E(q_i)}{T_{coll}}}.
\end{equation}
Here $T_{coll}$ is a parameter characterizing the width of the distribution (\ref{boltz}) in the space of deformation parameters.

The energy $E(q_i)$ in (\ref{boltz}) is the sum of the liquid-drop deformation
energy $E_{LDM}$ and of the shell correction $\delta E$.
As in any scission-point model, it is E$_{def}$ along the scission line that determines the main features of the calculated distribution: the most probable masses and the relative intensities of the fission modes.

The distribution (\ref{boltz}) is a basic assumption of the model.
In fact, in actual calculations,
this assumption is made only at scission and only in the mass asymmetry and
other modes which are normal to the fission mode.
No assumption is made concerning the motion in the fission direction which means that
we can consider at scission non-zero translational kinetic energy.

There is only one parameter T$_{coll}$.
It has no physical meaning. Besides controlling the overall width of the distribution,
it accounts also for the finite experimental mass resolution.
It was always fitted.
In \cite{spm} T$_{coll}$ was close to $1 \mev$. In \cite{CIOT} was chosen $1.5 \mev$
In the present calculations we use $T_{coll}$=2 MeV, from a previous comparison with
experimental distributions in No and Rf isotopes \cite{cluster16}

\section{The Cassini shape parametrization}
\label{potential}

To correctly define the nuclear shape at the scission point and calculate the potential energy $E(q_i)$ one would need, first of all, a flexible shape parametrization.

In the present work we use the shape parametrization based on Cassinian ovaloids introduced in \cite{pash71,pash88}. The axially symmetric Cassinian ovaloids are defined by the rotation of the curve
\begin{equation}\label{cassini}
\bar\rho(\bar z)=[\sqrt{1+4\eps \bar z^2}-\bar z^2-\eps]^{1/2}
\end{equation}
around $z$ axes. Here $\bar\rho\equiv\rho/R_0, \bar z\equiv z/R_0)$, where $R_0$ is the radius of the spherical nucleus,
and $\eps$ is the deformation parameter that fixes the elongation of the shape.
\begin{figure}[htp]
\includegraphics[width=0.4\textwidth]{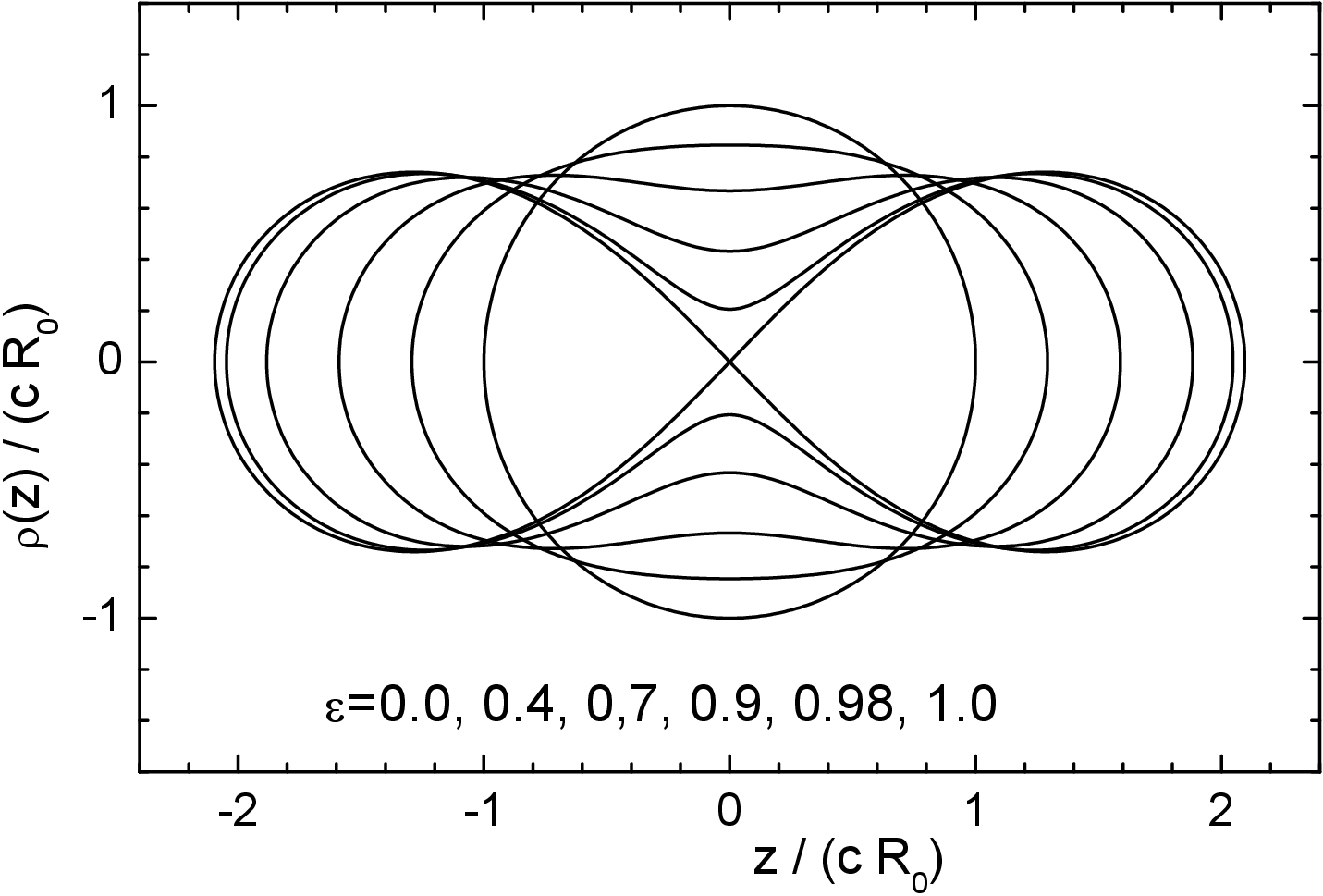}
\centering
\caption{The Cassinian ovals (\ref{cassini}) for few values of the deformation parameter $\eps$, indicated in the figure.}
\label{cassin}
\end{figure}
The value $\eps=0$ corresponds to a sphere. For $0<\eps<0.4$ the shape (\ref{cassini}) resembles very much that of spheroids with the ratio of semiaxes given by
\begin{equation}\label{ratio}
\frac{\rm{shorter\,\,axes}}{\rm{longer\,\, axes}}=\frac{1-2\eps/3}{1+\eps/3}\,.
\end{equation}
At $\eps\approx0.5$ the neck appears and at $\eps=1$ the neck becomes zero, see. Fig. \ref{cassin}.

In ref. \cite{rabotnov} it was shown that Cassinian ovals are a convenient single-parameter family of curves that surprisingly well approximate the nuclear shape at the saddle point.

For more complicated shapes, in \cite{pash71} the cylindrical coordinates $(\rho,z)$ were expressed in terms of the lemniscate coordinates $(R,x)$
\begin{eqnarray}\label{Rx}
\rho&=&[(p(x)-R^2(2x^2-1)-s]^{1/2}/\sqrt{2}\,,\\
z&=&[(p(x)+R^2(2x^2-1)+s]^{1/2}/\sqrt{2}\,,\\\nonumber
\text{with}\,p(x)&\equiv& [R^4+2sR^2(2x^2-1)+s^2]^{1/2}\,.\nonumber
\end{eqnarray}
$0\leq R<\infty$ and x is confined by $-1\leq x \leq 1$. Here $s\equiv\eps R_0^2$.
In lemniscate coordinates the profile function is given by the equation $R=R(x)$. For $R=R_0=const$ the shape coincides with
the Cassinian ovals (\ref{cassini}). For more complicated shapes one has to introduce the dependence of $R$ on $x$. In \cite{pash71} $R(x)$ was represented as the series in Legendre polynomials $P_n(x)$
\begin{equation}\label{alphas}
R(x)=(R_0/c)(1+\sum_n\alpha_n P_n(x)).
\end{equation}
The coefficients $\alpha_n$ are considered as the deformation parameters, additional to $\eps$. In (\ref{alphas}) and Fig. 1 the constant $c$ is a scaling factor that guarantees the conservation of volume inside of Cassinian ovaloid under the variation of deformation parameters.

Near the scission point, it is better to use an independent parameter $\alpha$, instead of $\eps$, see \cite{pash88},
\begin{equation}\label{alpha}
\alpha=(z_L^2+z_R^2-2\rho_{neck}^2)/(z_L^2+z_R^2+2\rho_{neck}^2),
\end{equation}
where $z_L (z_R)$ are the minimal(maximal) values of the coordinate $z$.
The advantage of $\alpha$ is that at $\alpha=1$ the neck radius turns into zero independently of all other deformation parameters $\alpha_n$.

The set of deformation parameters $\alpha, \alpha_n$ and the equations (\ref{Rx}) supply the expression for the profile function $\rho(z)$ in a parametric form, $\rho=\rho(R(x),x), z=z(R(x),x)$, for $-1\leq x\leq 1$.

\begin{figure}[htp]
\includegraphics[width=0.45\textwidth]{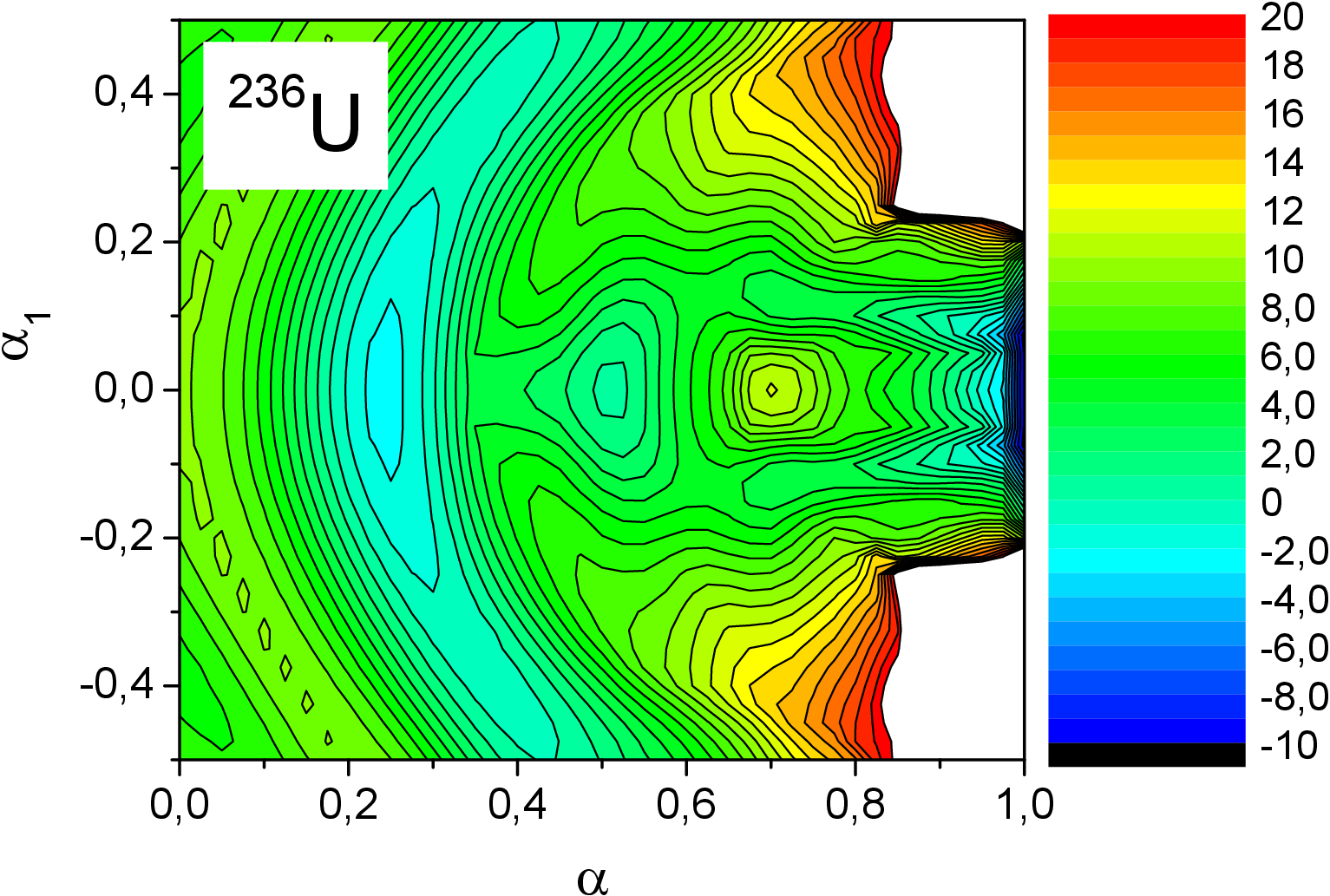}
\centering
\caption{(color online) The potential energy  (\protect\ref{edef},\ref{deltae}) of $^{236}$U  minimized concerning $\alpha_4$ at fixed $\alpha$ and $\alpha_1$ deformation parameters.}
\label{edef236}
\end{figure}

Already with three deformation parameters $\alpha, \alpha_1$ and $\alpha_4$ the Cassinian shape parametrization presents a quite reasonable family of shapes for the fission process. The deformation energy (\ref{edef},\ref{deltae}) minimized with respect
to $\alpha_4$ is shown in Fig. \ref{edef236} for $^{236}$U.  The ground state,
the second minimum and the mass-asymmetric saddle are seen quite well there.

\section{The potential energy along the scission line}
\label{transp}
In the present work, we define the scission shape by the fixed value of $\alpha$, $\alpha=0.98$, and take into account five additional deformation parameters,  $\alpha_{1},\alpha_{2},\alpha_{3},\alpha_4$ and $\alpha_{5}$,
that vary within the limits: $-0.5\leq\alpha_{1}\leq 0.5,\,\Delta\alpha_1=0.025, \,0\leq\alpha_{2}\leq 1,\,\Delta\alpha_2=0.1, \,-0.75\leq\alpha_{3}\leq 0.75,\,\Delta\alpha_3=0.05,  \,-0.3\leq\alpha_{4}\leq 0.3, \,\Delta\alpha_4=0.02$, and $-0.25\leq\alpha_{5}\leq 0.25, \,\Delta\alpha_5=0.05$.
\begin{figure}[htp]
\includegraphics[width=0.45\textwidth]{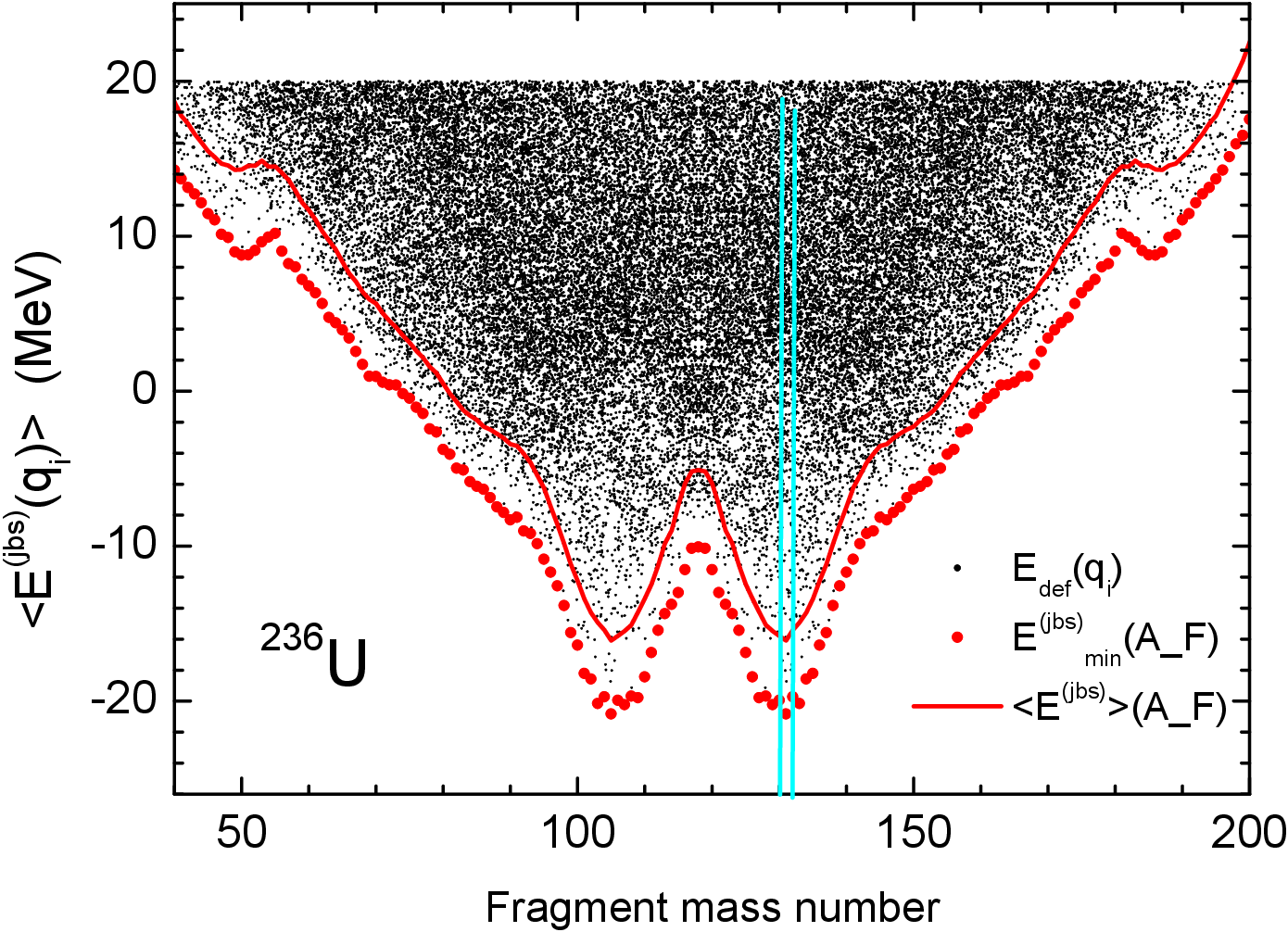}
\centering
\caption{(color online) The deformation energy (\ref{edef}) (black points) and the mean value $E^{(jbs)}(q_i)$ (\ref{Ejbs}) (red line) as a function of the fragment mass number.}
\label{poten}
\end{figure}

With these shape parameterizations, we calculate the potential energy of
deformation using the microscopic - macroscopic approach \cite{Strutinsky,brdapa}:
\begin{equation}\label{edef}
E_{def} (q_i)=E_{def}^{LD} (q_i) + \delta E (q_i)\,,
\end{equation}
with
\begin{equation}\label{deltae}
\delta E(q_i)= \sum_{n,p}[\delta E_{shell}^{(n,p)}(q_i)+\delta E_{pair}^{(n,p)}(q_i)].
\end{equation}
The summation in (\ref{deltae}) is carried out over the protons
($p$) and neutrons ($n$).
The $E_{def}^{LD}$ in (\ref{edef}) is the macroscopic liquid-drop energy
and $\delta E$ contains the microscopic shell and pairing corrections calculated
with the  Woods-Saxon potential \cite{pash71} for the  shape given by parameters $\{q_i\}$.

In what follows we will calculate the fission fragment mass distribution as a function of the mass asymmetry, or the fragment mass number.
Here one should keep in mind that in Cassini parametrization the mass asymmetry is not an independent deformation parameter. At $\alpha$=0.98 each point $\{q_i\}\equiv \{\alpha_{1i},\alpha_{2i},\alpha_{3i},\alpha_{4i}, \alpha_{5i}\}$ in 5-dimensional deformation space has its own mass asymmetry $\delta_i\equiv (V_L(i)-V_R(i))/(V_L(i)+V_R(i))$, where $V_L(i)$ and $V_R(i)$ are the volumes to the left and right from the plane $z=0$. The point $z=0$ coincides with or is very close to the smallest neck radius.

The fragment mass number is expressed in terms of $\delta_i$ as
\begin{equation}\label{AFi}
A_{Fi}=0.5 A (1+\delta_i),
\end{equation}
where $A$ is the mass number of the fissioning nucleus.

The deformation energy (\ref{edef}) as a function of the fragment mass number $A_{Fi}$ is shown by black dots in Fig. \ref{poten}.
The red dots mark the minimal value of the energy (\ref{edef}) at fixed $A_{F}$. The minimization is carried out over all $A_{Fi}$
closest to the integer number $A_{F}$. Here one should note that in multidimensional deformation space, there are a lot of local minima.
As $A_{F}$ changes, the minimal deformation energy jumps from one local minimum to another. The minimal energy shape is not a smooth function of $A_F$. If $A_F$ is changed a little, the minimal shape could be in another local well
far away from the previous in deformation space.
 To get the smooth dependence of the shape and deformation energy, one could consider the mean value $\bra E^{(jbs)}(A_F)\ket$, with the averaging being done with the canonical distribution  (\ref{boltz}).

The mean value $\bra E^{(jbs)}(A_F)\ket$ of the deformation energy is defined then as the contribution from the points in deformation space whose fragments mass $A_{Fi}$  are closest to  $A_F$, $A_F-1/2\leq A_Fi\leq A_F+1/2$,
\begin{equation}\label{Ejbs}
\bra E^{(jbs)}\ket(A_F)=\frac{\sum_{|A_{Fi}-A_F|\leq 1/2}E_{def}(q_i)e^{-E_{def}(q_i)/T_{coll}}}
{\sum_{|A_{Fi}-A_F|\leq 1/2}e^{-E_{def}(q_i)/T_{coll}}}
\end{equation}

The potential energy (\ref{Ejbs}) for $^{236}$U is shown in Fig. \ref{poten} by the red
line. For $A_F$=132, for instance, the summation in (\ref{Ejbs}) is carried out over
all the points between the two vertical lines in Fig. \ref{poten}.
Only a small part of the black points (about 3$\%$) is included in Fig. \ref{poten}. Otherwise, the size of Fig. \ref{poten} would become unmanageable.
For $A_F$=132 the summation in (\ref{Ejbs}) is carried out over all the points between
the two vertical lines in Fig. \ref{poten}.

\begin{figure}[htp]
\includegraphics[width=0.45\textwidth]{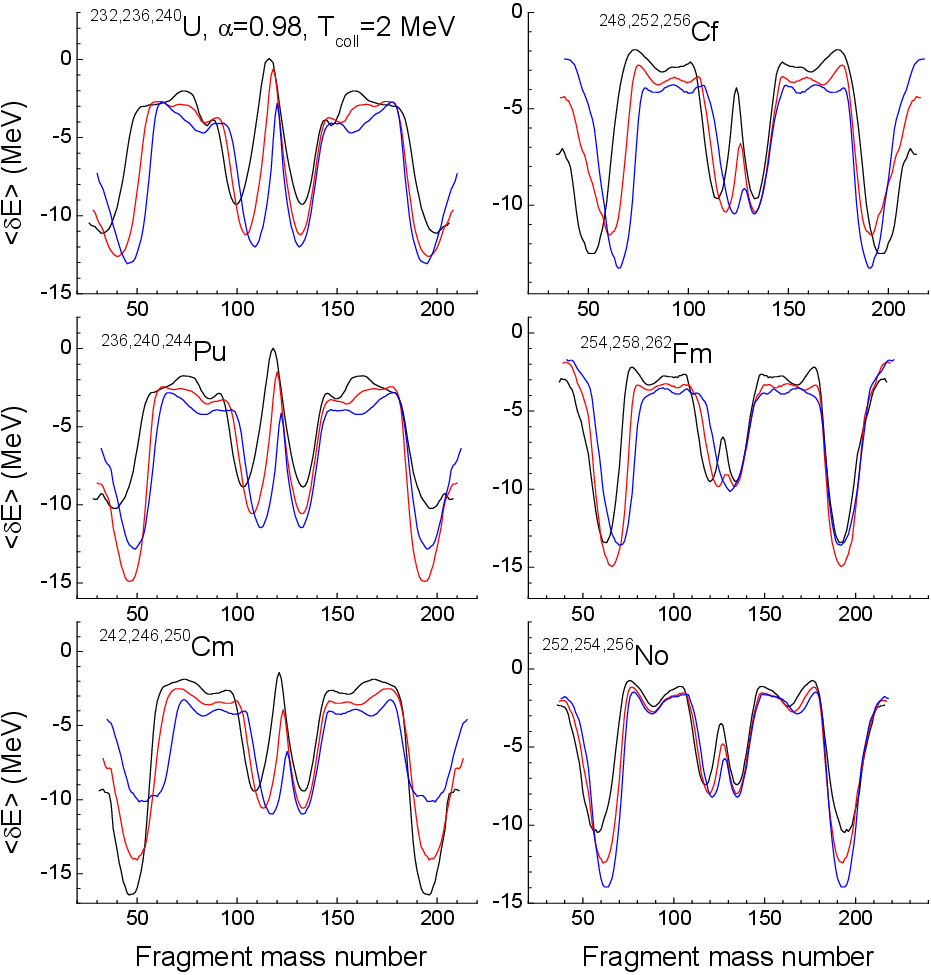}
\centering
\caption{(color online) The shell correction (\ref{deltae}) for isotopes of the few nuclei along the scission line as a function of the fragment mass number.}
\label{shelco}
\end{figure}
The interesting feature of Fig. \ref{poten} is that besides the known minimum at $A_F$ close to 132,  one can see an additional super-asymmetric minimum at $A_F$ close to  190. This super-asymmetric minimum is much higher in energy as compared with the standard mass-asymmetric minimum. So, the yield of the mass asymmetric peak will be about 6 orders of magnitude smaller compared with the standard peak, see the top right part of Fig. \ref{yields}.

One can also observe that the energy of SAF isomer is positive which means that it
cannot be populated by spontaneous fission. One needs at least 15 MeV to reach it.
Another consequence of the high energy of the shoulder is that SAF will always occur at
about 35 MeV lower excitation, i.e., towards the cold fission limit. It is therefore not surprising that the even-odd effect increases strongly in the region of
super-asymmetric nuclear charge \cite{KarlHeinz,Rochman}.

The super-asymmetric minimum at $A_F$=190 is produced by a large shell correction. The dependence of the shell corrections on the mass asymmetry for several nuclei is presented in Fig. \ref{shelco}. For all shown here nuclei the shell corrections for the super-asymmetric shape are (much) larger than for the standard asymmetric shape. They are therefore expected to better resist at high excitation energies.

\begin{figure}[ht]
\includegraphics[width=0.48\textwidth]{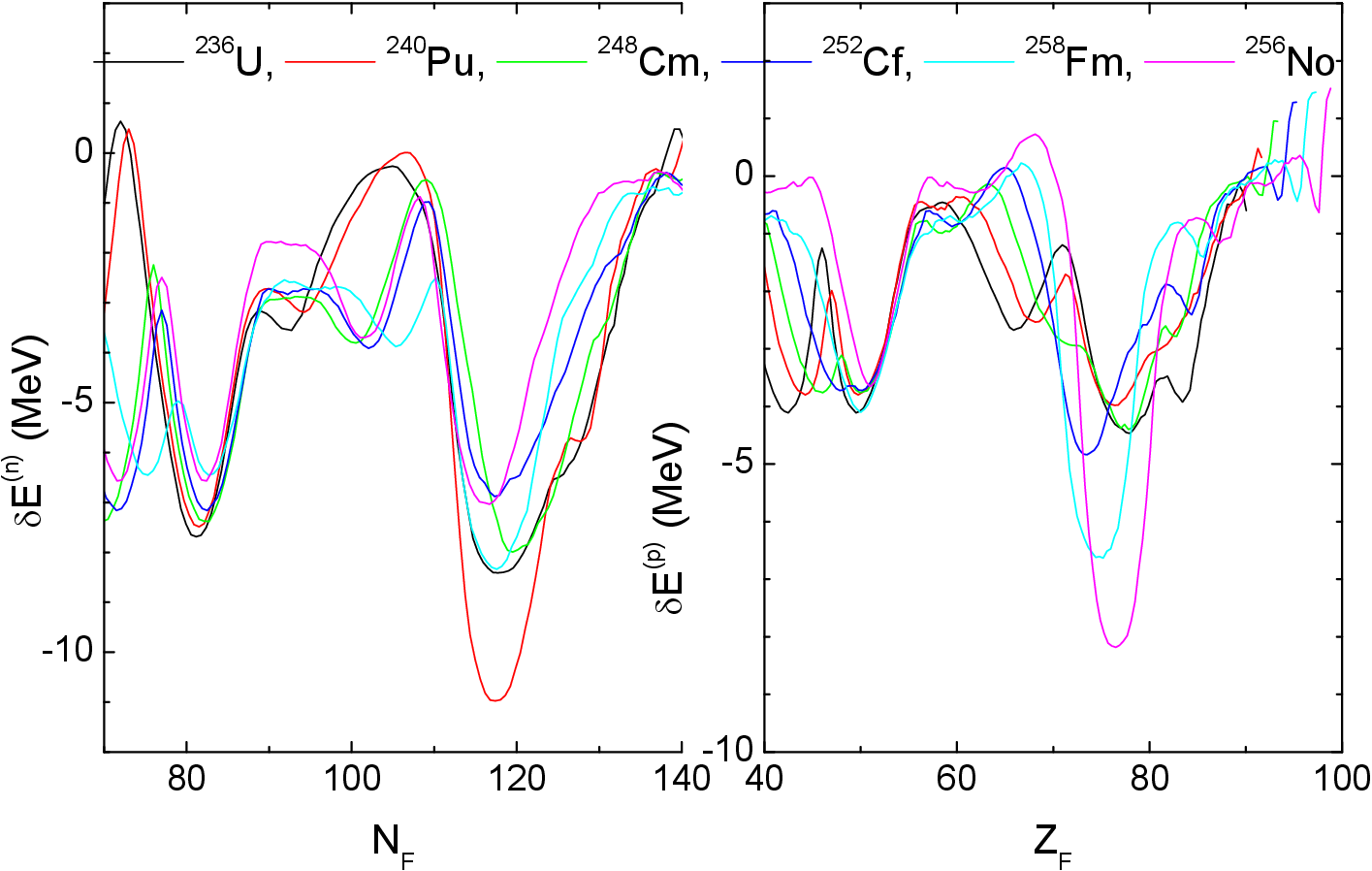}
\centering
\caption{(color online) The neutron and proton shell corrections (\ref{deltae})
as a function of the mass asymmetry (the number of neutrons $A_N$ or protons $A_Z$ in heavy fragment).}
\label{shcos}
\end{figure}

In Fig. \ref{shelco} the shell correction of the whole nucleus is plotted as a function of the mass number of the heavy fragment $A_F$.
But shell effects are associated with specific values of N and Z, rather than A.
To find out what those 'magic' numbers are, we show in Fig \ref{shcos}
the shell corrections for neutrons and
protons as a function of mass asymmetry (the number of
neutrons $A_N$ or protons $A_Z$ in heavy fragment),
$N_F=A_F N_{CN}/A_{CN}, Z_F=A_F Z_{CN}/A_{CN}$,
where $N_{CN}, Z_{CN}$ and $A_{CN}$ are the neutron,
proton and mass numbers of the fissioning nucleus.
At small mass asymmetry one
clearly sees the magic numbers $N=82$ for neutrons and $Z=50$
for protons.
These magic numbers form the doubly magic nucleus $^{132}$Sn
that contribute a lot to the standard asymmetric fission of actinides.
So, the SPM results are close to the experiments for the standard asymmetric fission.
Besides, one can see another pair of magic numbers N close to 118
for neutrons and Z close to 76 for protons. These numbers
contribute to the mass number $A_F\approx$ 190 that we got for super
asymmetric fission. The shift 194 $\rightarrow$ 190 is due to the slope of the
LD potential. Thus, the heavy fragments in super asymmetric
fission are close to $^{208}$Pb. The difference between $A_F$=190 and
$^{208}$Pb can be related to the effect of light dumbbell fragments on
the total shell correction.
\section{The fission fragments mass distribution (FFMD)}
\label{results}
The FFMD is defined as the normalized to 200\% sum of the distributions (\ref {boltz}) over all deformation points $q_i$  with the same mass asymmetry $\delta$. In Cassinian shape parametrization $\delta$ is not an independent parameter and the values $\delta_i$ of individual trajectory may not coincide with the fixed $\delta$.
It makes sense then to account for all contributions with the mass asymmetry $\delta_i$ close to $\delta$, i.e., to calculate the mean value,
\begin{eqnarray}\label{theyield}
\bra Y(A_F)\ket=200 \sum_{|A_{Fi}-A_F|\leq 1/2}e^{-E_{def}(q_i)/T_{coll}}/N,\\
\text{with}\quad N\equiv\sum_{A_F}\sum_{|A_{Fi}-A_F|\leq 1/2}e^{-E_{def}(q_i)/T_{coll}}.\nonumber
\end{eqnarray}

\begin{figure}[htb]
\centering
\includegraphics[width=0.48\textwidth]{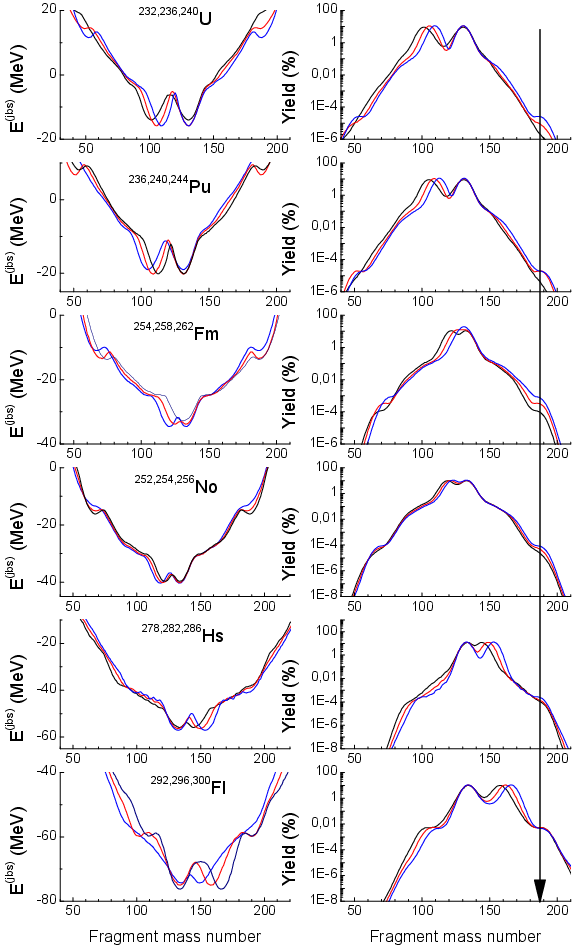}
\caption{(color online) Left: The mean value $E^{(jbs)}(q_i)$ (\ref{Ejbs}) of the potential energy just before the scission as a function of the fragment mass number. Right: The calculated FFMDs for the isotopes of a few nuclei from U to Fl as a function of the fragment mass number. }
\label{yields}
\end{figure}

The calculated yields of fission fragments for the isotopes of a few nuclei from uranium to flerovium are shown in the right part of Fig. \ref{yields}.
One can see the super mass-asymmetric peak at $A_F$=190 present in the FFMD of all considered nuclei. The relative magnitude of the super mass-asymmetric peak is of the order of $10^{-6}$ for U isotopes and close to $10^{-2}$ for Fl. Thus, in most of
these nuclei, the super mass-asymmetric mode could be observed experimentally.
There is a major difference between our results and the Lohengrin data \cite{Rochman}.
Here the SAF occurs at a constant heavy-fragment mass (A$_F$=190) and not at a constant
light-fragment mass (A$_F$=70). The extra stability comes from a large
shell effect in dumbbell nuclei when the "spherical" heavy fragment has the mass
$\approx$ 190. See Fig.4.
So the shell stabilization is associated with the heavy fragment group.

Comparing the left and right parts of Fig. \ref{yields} one sees the surprising similarity of the potential energy just before the scission and the fission fragments mass distribution. This similarity is a consequence of the definition (\ref{theyield}). For each trajectory, the yield is simply proportional to $\exp{[-E^{(jbs)}/T]}$.

\begin{figure}[ht]
\includegraphics[width=0.8\columnwidth]{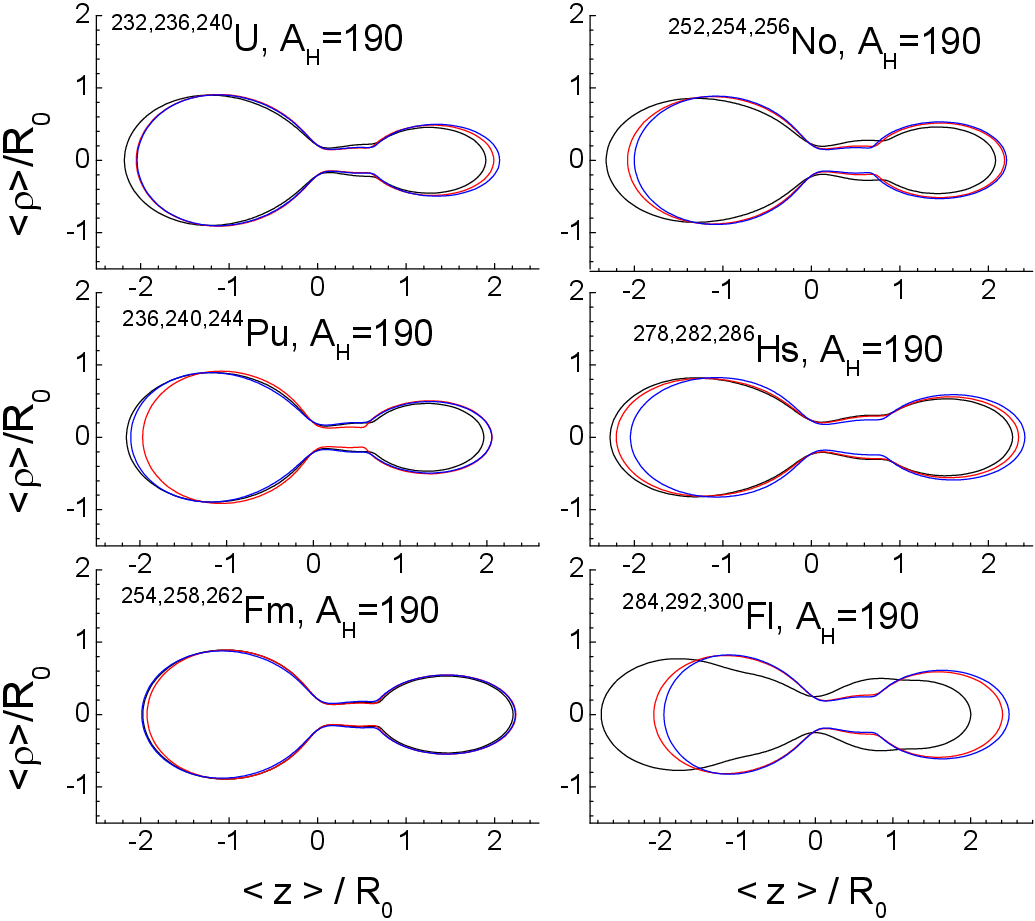}
\caption{(color online)The average shapes of fissioning nuclei at the scission point at the super mass-asymmetric maximum of FFMD. The colors, black, red, and blue, change from smaller to larger numbers of neutrons.
.}
\label{shapes}
\end{figure}

In Fig. \ref{shapes}, we show the average scission shapes of fissioning nuclei at the super mass-asymmetric maximum of FFMD.

Here we should mention that the definition of average scission shapes requires special care. For each $q_i$ the scission shapes look very different, even the elongation of the shape $z_R(i)-z_L(i)$ is not the same.

We calculate the profile function $\rho(z)$ at the set of the points $z_k$ that are related to the variable $-1\leq x_k \leq 1$ by the transformation (\ref{Rx}).
The points $x_k$ are the abscissas of the Gaussian integration method.  In the calculations, we used 258 points $x_k$ to ensure a high accuracy of calculations of integrals for the volume, position of the center-of-mass, or multiple moments of the density distributions. The set of points $x_k$ is the same for any deformation of the shape. Then we average in deformation the points $z^{(i)}$ and $\rho^{(i)}$ related to the same value of variable $x_k$,
\begin{eqnarray}\label{scishape}
\bra \rho_k\ket=\sum_i \rho^{(i)}(x_k)e^{-E_{def}(q_i)/T_{coll}}/N\,,\\
\bra z_k\ket=\sum_i z^{(i)}(x_k)e^{-E_{def}(q_i)/T_{coll}}/N\,,\\\nonumber
\text{with}\quad N\equiv\sum_i e^{-E_{def}(q_i)/T_{coll}}\,.\nonumber
\end{eqnarray}
As one can see from Fig. \ref{yields}, the SAF scission shapes for all considered nuclei have the form of an asymmetric dumbbell. The reason for potential minima at very large mass asymmetries is the occurrence of these types of shapes. If one moves away from the relative minima the dumbbells disappear. Phenomenological models that cannot describe
dumbbell shapes cannot predict SAF. Even without $\alpha_5$, generalized Cassini ovals
can describe perfect dumbbells proving, once again,  their superiority at scission.

 A similar situation occurs at the absolute potential minima (i.e., at A$_F \approx$ 132)
 where pear-shaped fragments are the reason for these minima.
 For the important role of the octupole
 deformation in the formation of the fission fragments, see refs. \cite{CIO,SS-Nature}.
 This association of the minima of E$_{def}$ along the scission line with very
 peculiar nuclear shapes (dumbbells or pear shapes) is intriguing and cannot be a hazard.
 It may have implications on the scission process.

It is worth noting that the scission shape with the smallest second neck radius corresponds to the largest magnitude of the SAF peak in FFMD. They also correspond to
better dumbbells.

\section{The possibility of ternary fission}
\label{ternary}
In very heavy nuclei one may expect new phenomena that could not be observed in lighter nuclei. One example is the ternary fission. For actinide nuclei, ternary fission usually means light-charged-particle-accompanied binary fission. For the super-heavy nuclei one may expect fission in three fragments of comparable masses.

Looking at the dumbbell shapes in Fig.\ref{shapes} one could expect that after the
first neck ruptures (on the heavy fragment side), the second, slightly thicker, neck
will also break into two even lighter fragments. The amount of the mass between the two necks is shown in the table below.
\begin{table}[h]\label{table}
\caption{(color online) The mass (in amu) of the heavy ($A_H$) and  light ($A_L$) fragments in the super mass-asymmetric fission of few heavy and super-heavy nuclei and the amount of mass ($A_3$) in the neck region of light fragment.}
\begin{center}
\begin{tabular}{|l||c|c|c|c|c|c|c|r|} \hline
\,&$^{236}$U&$^{240}$Pu&$^{246}$Cm&$^{252}$Cf&$^{258}$Fm&$^{256}$No&$^{282}$Hs&$^{292}$Fl\\ \hline
$A_H$&190&190&190&190&190&190&190&190\\ \hline
$A_L$&46&50&56&62&68&66&92&102\\ \hline
$A_3$&2&2&4&3&2&2&10&6\\ \hline
\end{tabular}
\end{center}
\end{table}

The neck mass is almost constant except for Hs and Fl where it is slightly larger. The constancy of the neck mass $A_3$ is a consequence of the Rayleigh instability of the fluid cylinders that occurs at fixed value of $l_{neck}/r_{neck}$ and was already discussed in connection with the ternary fission in \cite{CSN}.

In order to decide whether the light deformed fragment will fission or will recover a
compact shape, we examine the slope of the multidimensional potential energy at the position of the
light fragment.

\begin{figure}[ht]
\includegraphics[width=0.8\columnwidth]{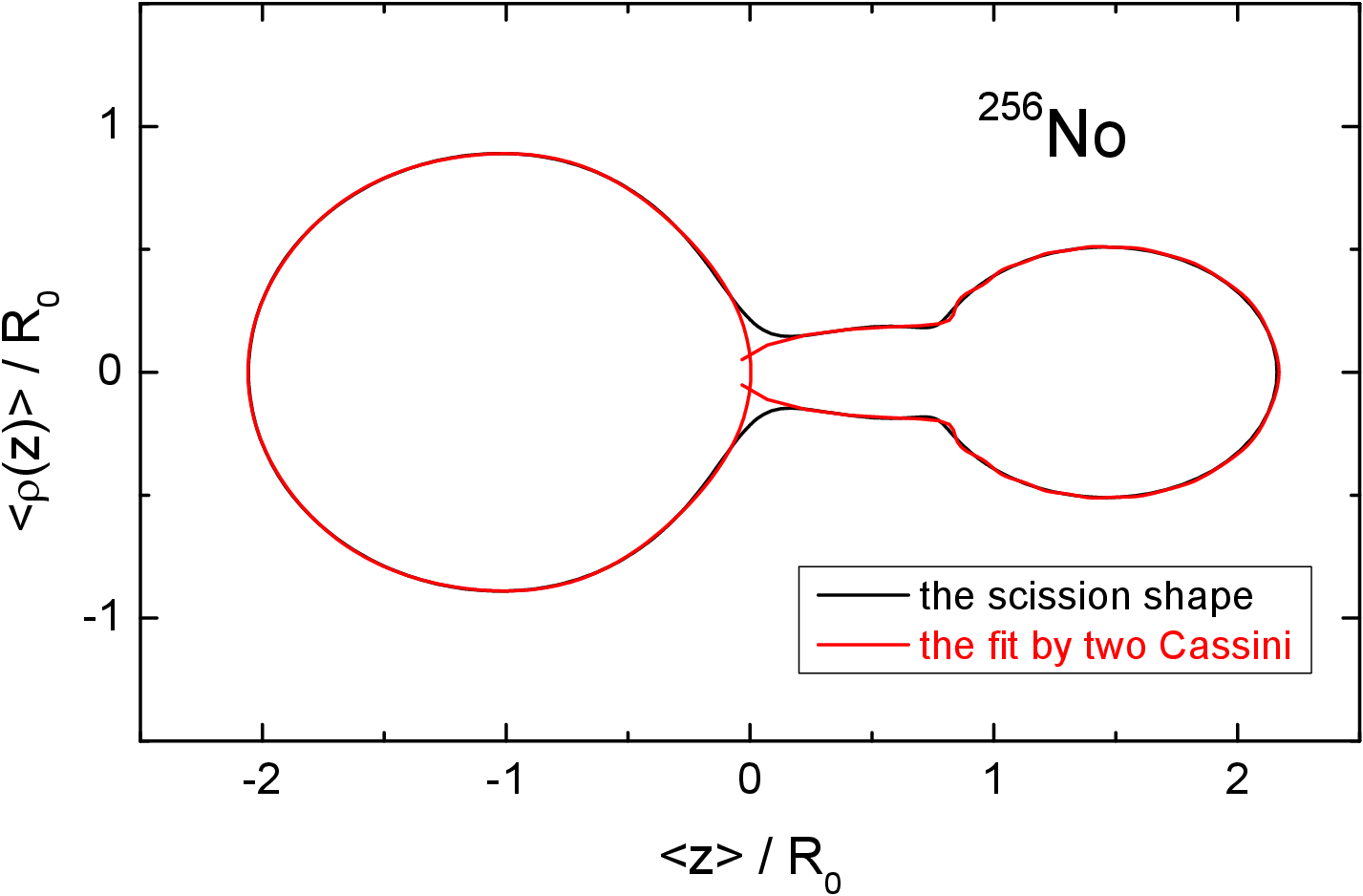}
\caption{(color online) The fit of the just-before-scission shape of $^{256}$No (black) by two separated fragments (red)
.}
\label{ternar}
\end{figure}

The fit of the light fragment, shown in Fig. \ref{ternar}, gives the values of all deformation parameters $\alpha_1 - \alpha_{16}$. Then we calculate the derivative of potential energy for each deformation parameter $\alpha_n$ and make a small step i.e., vary each deformation parameter by 0.01 in the direction of smaller deformation energy.

\begin{figure}[ht]
\includegraphics[width=0.9\columnwidth]{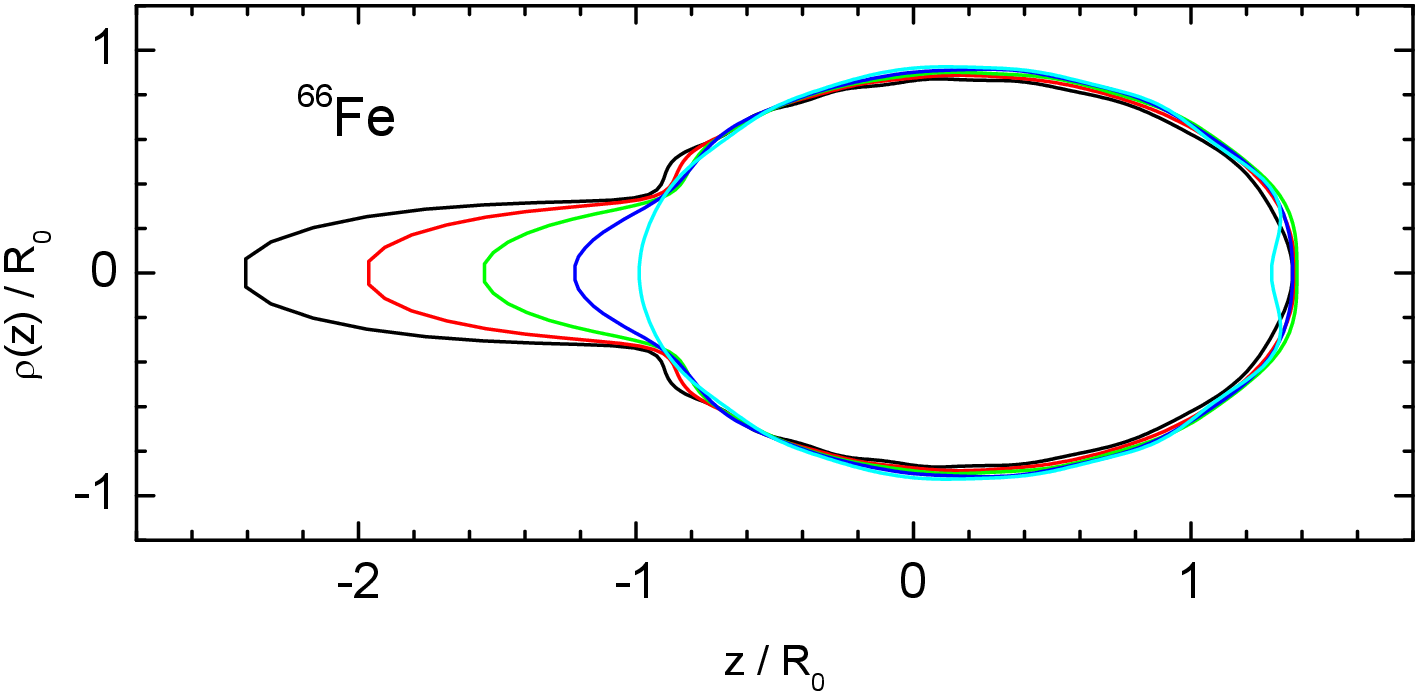}
\caption{(color online) The variation of the shape of light fragment along the valley of multi-dimensional potential energy. Red, green. blue and cyan colors mark to results of 3, 6, 9 and 12 steps towards the minimum of potential energy.}
\label{shapeAL}
\end{figure}

We repeat the procedure several times. The evolution of the shape after 3, 6, 9 and 12 steps is shown in Fig. \ref {shapeAL}. As one can see, the slope of potential energy drives the light fragment towards more compact shape. Thus, if there is no temperature in the system, the sequential ternary fission following SAF is not energetically
possible.


At this point we should remember that
the probability of sequential  ternary fission of $^{252}$Cf following the main asymmetric fission mode (A$_H \approx$ 132) was estimated in \cite{wada21} considering
a fixed temperature T=1 MeV.
The evolution of the shape of the light fragment (A$_H \approx$ 120) on a two-dimensional ($\alpha,A_F$) PES was calculated using the Metropolis method. It was found that, even if the light fragment at scission is below its saddle point, a small fraction (10$^{-6}$) of trajectories pass the saddle due to random forces. This rare sequential fission of the light fragment produces an even lighter fragment of mass A=69 which was associated with the "Ni bump" observed in collinear cluster tri-partition \cite{Pyatkov}.
At this moment we cannot exclude that the shoulder observed at Lohengrin \cite{Rochman}
has the same origin, i.e., sequential ternary fission.

\section{The total kinetic energy}
\label{kinetic}

Among the nuclei considered in the present study, $^{256}$No and $^{254}$No have the
best chance
to be synthesized in amounts large enough to observe their SAF mode.
One possibility would be a heavy ion reaction using $^{48}$Ca projectile on $^{208}$Pb
target at the SHE-Factory constructed at FLNR of JINR-Dubna.
The compound nuclei will have an
excitation energy around 20 MeV and will subsequently undergo fission, either directly
or after the emission of two neutrons.
The production cross-sections of these nuclei being of the order of
two microbarns, few 10$^{6}$ fission events could be collected.

\begin{figure}[ht]
\includegraphics[width=0.8\columnwidth]{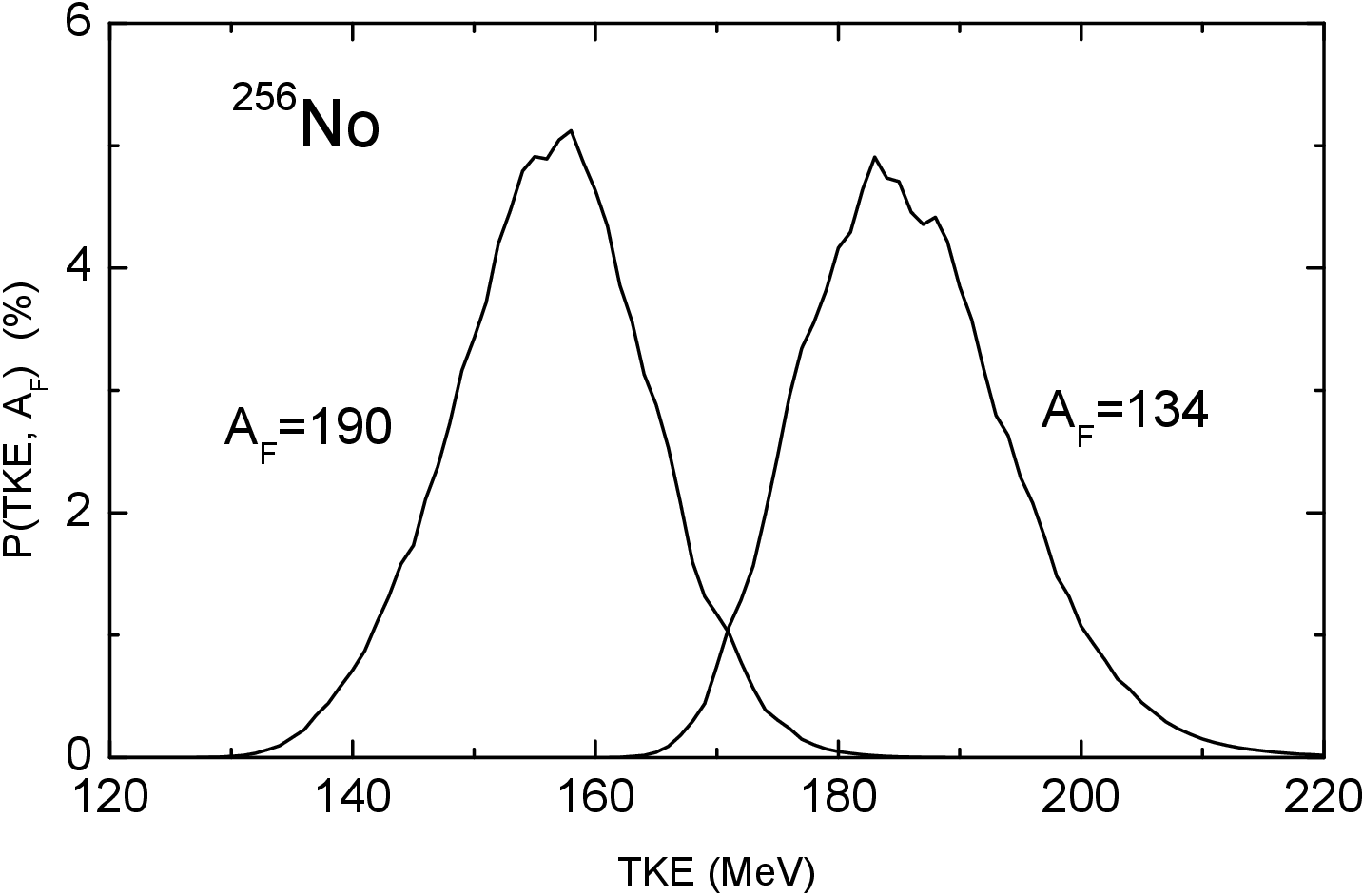}
\caption{The calculated TKE distributions $P(TKE, A_F)$ corresponding to narrow windows around the main mass peak
$A_F$=134 and the SAF peak $A_F$=190. }
\label{tkedistr}
\end{figure}

Envisaging this possibility we calculate in this section the expected TKE
distributions $P(TKE, A_F)$ corresponding to narrow windows ($\Delta A_F$=2) around the main mass peak and the
SAF peak, see Fig. \ref{tkedistr}. As one could expect, the average value of TKE at the main peak is larger
 than that at SAF peak. The width of distributions in both cases is approximately equal to 20 MeV.
 The information below will help the identification of these two fission modes.

The distributions $P(TKE, A_F)$ shown in Fig. \ref{tkedistr} is defined as
\begin{equation}\label{PTKE}
P(E, A_F)=100\sum_{|E_i-E|\leq 5 MeV}\sum_{|A_{Fi}-A_F|\leq 2}
e^{-\frac{E_{def}(q_i)}{T_{coll}}}/N
\end{equation}
with the normalization factor
\begin{equation}\label{NTKE}
N=\int_{0}^{\infty}dE\sum_{|E_i-E|\leq 5 MeV}\sum_{|A_{Fi}-A_F|\leq 2}
e^{-\frac{E_{def}(q_i)}{T_{coll}}}.
\end{equation}
Here $E$ stands for the total kinetic energy, $TKE$.

We define the TKE as the Coulomb interaction of the nascent fragments at $\alpha$=0.98
in the point-charge approximation since the prescission kinetic energy cannot be calculated within SPM.
The average value $E_{Coul}^{(int)}$ around a given fragment mass $A_F$
is
\begin{equation}\label{Ecoul_mean}
E_{Coul}^{(int)}(A_F)\equiv\sum_{|A_{Fi}-A_F|\leq \Delta A_F}
\frac{e^2Z_{Li}Z_{Hi}}{2R_{12}(q_i)}e^{-\frac{E_{def}(q_i)}{T_{coll}}}/N\,.
\end{equation}
$eZ_{Li} $ and $eZ_{Hi} $ are the charges of light and heavy fragments.
The averaging interval $\Delta A_F$ in present calculations was put equal to 2 amu.

One can see from (\ref{Ecoul_mean})
that the Coulomb interaction energy depends crucially on the mean value of the distance between centers of mass of fragments $<R_{12}^{(crit)}(\delta)>$.
The small $<R_{12}^{(crit)}(\delta)>$ leads to the large value of $<E_{Coul}^{(int)}(\delta)>$ and, on the contrary, the large $<R_{12}^{(crit)}(\delta)>$ leads to the small value of $<E_{Coul}^{(int)}(\delta)>$.

\begin{figure}[ht]
\includegraphics[width=0.8\columnwidth]{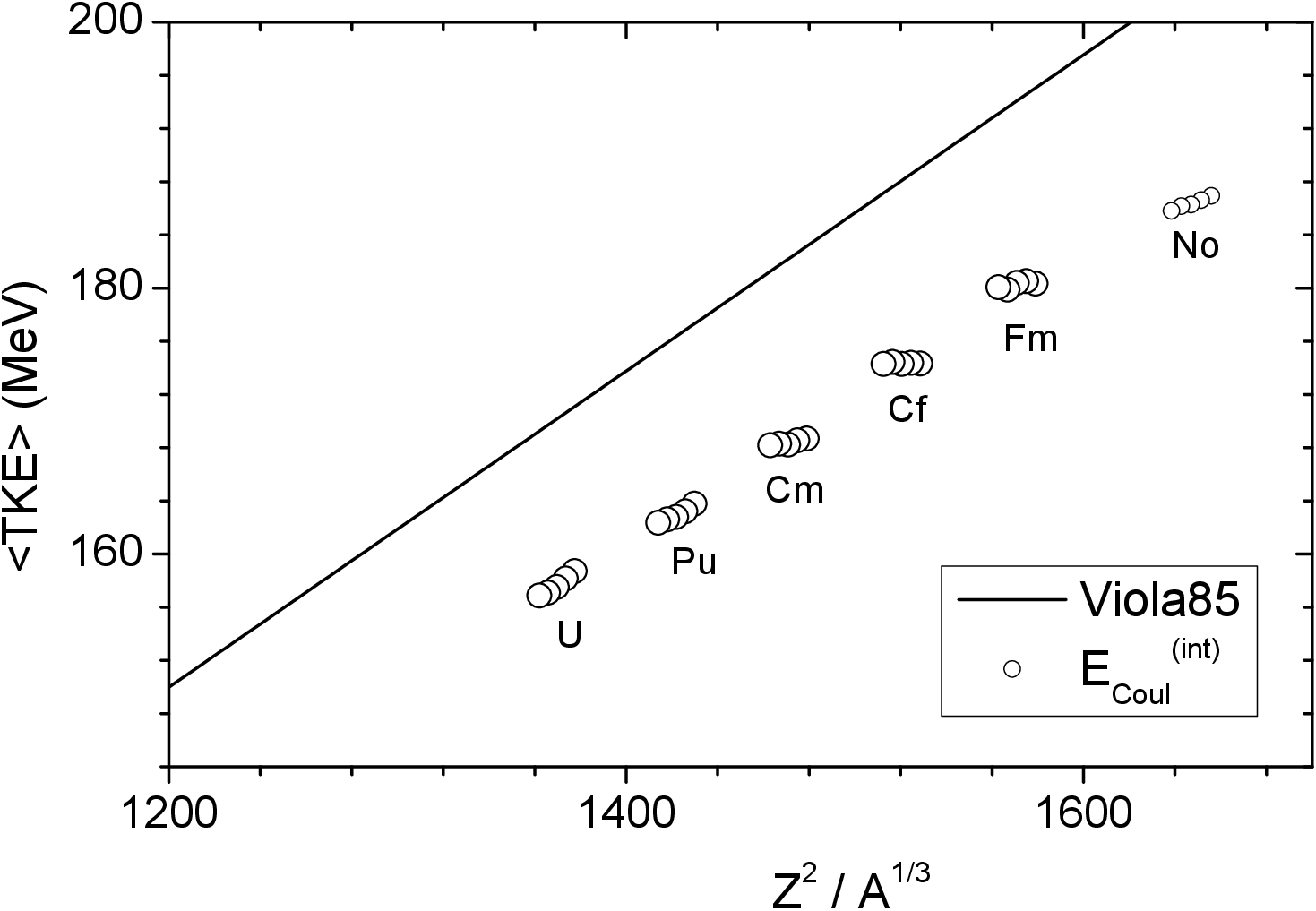}
\caption{Coulomb repulsion energy (17) and Viola systematics (\ref{viola}) for the TKE.}
\label{tottke}
\end{figure}

The comparison of the calculated values of $TKE$ averaged over all fragment masses with Viola systematics \cite{viola1985}
\begin{equation}\label{viola}
\bra E_K\ket= (0.1189\pm 0.0011)Z^2/A^{1/3} + 7.3(\pm 1.5)\rm {MeV}
\end{equation}
is show in Fig. \ref{tottke}.

As expected, the calculated $<$TKE$>$ underestimates the systematics  for all studied systems
and the neglected prescission contribution slightly increases with fissility (from 12
to 16 MeV).

\section{Summary}
\label{suma}
An improved scission-point model is used to calculate the fragment yields
over a large domain of fragment masses for several fissioning nuclei.
Isotopes of uranium, plutonium, fermium, nobelium, hassium and flerovium are studied.
In addition to the standard peak at
$A_F$=132, a pronounced shoulder is found at large mass asymmetries ($A_F$=190) for
all these nuclei. The probability of this SAF mode relative to the standard mode is
increasing with the charge and the mass of the system from 10$^{-6}$ in U
to 10$^{-2}$ in Fl.

The extra stability of this rare mode is due to a very large
shell effect when the heavy fragment has a mass 190 amu. The corresponding scission
shape turns out to be an asymmetric dumbbell with an almost spherical heavy fragment and a
very deformed light fragment.
This association of a peculiar shape with a minimum of potential energy at
scission is valid also at the most probable mass division.
There, the octupole-deformed,
pear-shaped fragments are responsible for the absolute minimum \cite{CIO,SS-Nature}.

The neck on the heavy fragment side breaks first since it is smaller.
The stability against a second
neck rupture is investigated by looking at the slope of the multidimensional
potential energy at the position of the light fragment. If there
is no temperature in the system, the slope drives the fragment towards a compact shape.

Finally, the TKE distributions corresponding to the most probable and the super-asymmetric
divisions are calculated for $^{256}$No for which SAF is likely to be measured.

{\bf Acknowledgments.}
The authors would like to express their gratitude to Acad. Yu.Ts. Oganessian for suggesting this study and numerous discussions and to Dr. C. Schmitt for valuable comments and suggestions.


\begin{thebibliography}{99}
\bibitem{HS} O. Hahn, F. Strassmann, Naturwissenschaften \textbf {27}, 11 (1939).
\bibitem{BW} N. Bohr, J.A. Wheeler, Phys. Rev \textbf {56}, 426 (1939).
\bibitem{Weizsaker} C.F. v. Weizsaker, Z. Physik. \textbf {96}, 431 (1935).
\bibitem{Strutinsky} V.M. Strutinsky, Nucl. Phys. {\bf 3}, 449 (1966); Nucl. Phys. {\bf A95}, 420 (1967); Nucl. Phys. {\bf A122}, 1 (1968).
\bibitem{VH} R. Vandenbosch, J.R. Huizenga, Nuclear Fission, Academic Press, New York (1973).
\bibitem{Sandulescu1978} A. Sandulescu, H.-J. Lustig, J. Hahn, and W. Greiner, J . Phys. G: Nucl. Phys. \textbf {4}, 279 (1978).
\bibitem{Lustig1980} H.J. Lustig, J.A. Maruhn, W. Greiner,  J. Phys. G: Nucl. Phys. \textbf {6}, 25 (1980).
\bibitem{our17}C. Ishizuka, X. Zhang, M.D. Usang, F. A. Ivanyuk, and S. Chiba, Phys. Rev. C {\bf 101}, 011601(R) (2020).
\bibitem{pasha}P.V. Kostryukov, A. Dobrowolski, B. Nerlo-Pomorska, M. Warda, Z.G. Xiao,  Y.J. Chen, L.L. Liu, J.L. Tian, and K. Pomorski, Chin. Phys. C \textbf{45}, 124108 (2021).
\bibitem{BSM}M. Albertsson, B.G. Carlsson, T. D\"{o}ssing, P. M\"{o}ller, J. Randrup, S. Aberg, Eur. Phys. Journ. A, 56:46 (2020).
\bibitem{poenaru2018}D.N. Poenaru, and R.A. Gherghescu, Phys. Rev. C \textbf{97}, 044621 (2018).
\bibitem{warda2018}M. Warda, A. Zdeb, and L.M. Robledo, Phys. Rev. C \textbf{98}, 041602(R) (2018).
\bibitem{cluster}Z. Matheson, S.A. Giuliani, W. Nazarewicz, J. Sadhukhan, and N. Schunck, Phys. Rev. C \textbf{99}, 041304(R) (2019).
\bibitem{Rao} V. Rao, V. Bhargava, S. Marathe, S. Sahakundu, R. Iyer, Phys. Rev. C \textbf {9}, 1506 (1974).
\bibitem{knyaz} G.N. Knyazheva, S.N. Dmitriev, M.G. Itkis, I.M. Itkis, E.M. Kozulin, T.
Loktev, Yu.Ts. Oganessian, Proceedings of the  International Symposium on
Exotic Nuclei (EXON 2012) Vladivostok, 1 - 6 October 2012, Eds. Yu.E.
Penionzhkevich, Yu.G. Sobolev. World Scientific, Singapore, ISBN
978-981-4508-85-8, p.167-174, 2013.
\bibitem{Rochman} D. Rochman I. Tsekhanovich, F. Gönnenwein, V. Sokolov, F. Storrer, G. Simpson, O. Serot,
Nucl. Phys. A \textbf {735}, 3 (2004).
\bibitem{Ni68}  R. Broda, B. Fornal, W. Krlas, T. Pawat, D. Bazzacco, S. Lunardi, C. Rossi-Alvarez, R. Menegazzo, G. de Angelis, P. Bednarczyk, J. Rico, D. De Acuña, P.J. Daly, R.H. Mayer, M. Sferrazza, H. Grawe, K.H. Maier, and R. Schubart, Phys. Rev. Lett. \textbf{74}, 868 (1995).
\bibitem{Ni68a} O. Sorlin, S. Leenhardt, C. Donzaud, J. Duprat, F. Azaiez, F. Nowacki, H. Grawe, Zs. Dombradi, F. Amorini,
A. Astier, D. Baiborodin, M. Belleguic, C. Borcea, C. Bourgeois, D.M. Cullen, Z. Dlouhy, E. Dragulescu,
M. Gorska, S. Grevy, D. Guillemaud-Mueller, G. Hagemann, B. Herskind, J. Kiener, R. Lemmon,
M. Lewitowicz, S.M. Lukyanov, P. Mayet, F. de Oliveira Santos, D. Pantalica, Yu.-E. Penionzhkevich,
F. Pougheon, A. Poves, N. Redon, M.G. Saint-Laurent, J.A. Scarpaci, G. Sletten, M. Stanoiu, O. Tarasov,
Ch. Theisen, Phys. Rev. Lett. \textbf{88}, 092501 (2002).
\bibitem{JEF2} The JEF-2.2 nuclear data library, NEA data bank, (2000).
\bibitem{Wahl} A. Wahl, in: J.W. Behrens, A.D. Carlson (Eds.), 50 Years with Nuclear Fission, vol. 2, American Nuclear Society, 525 (1989).
\bibitem {spm} B.D. Wilkins, E.P. Steinberg, and R.R. Chasman, Phys. Rev. C {\bf 14}, 1832 (1976).
\bibitem {CIOT} N. Carjan, F.A. Ivanyuk, Yu. Oganessian and G. Ter-Akopian, Nucl. Phys. A \textbf{942}, 97 (2015).
\bibitem{stlapo} V.M. Strutinsky, N.Y. Lyashchenko, and N.A. Popov, Nucl. Phys. \textbf{46}, 659 (1963).
\bibitem{micro-spm}J.-F. Lemaitre, S. Goriely, S. Hilaire, and J.-L. Sida, Phys. Rev. C {\bf 99}, 034612 (2019).
\bibitem{cluster16} N. Carjan, F. Ivanyuk, Yu.Ts. Oganessian, Journal of Physics: Conf. Series \textbf{863}, 012044 (2017).
\bibitem{CIO} N. Carjan, F.A. Ivanyuk, Yu. Ts. Oganessian, Nucl. Phys. A \textbf{968}, 453 (2017).
\bibitem{carjan2019} N. Carjan, F.A. Ivanyuk, Yu. Ts. Oganessian, Phys. Rev. C \textbf{99}, 064606 (2019).
\bibitem{pash71} V.V. Pashkevich, Nucl. Phys. A \textbf{169}, 275 (1971).
\bibitem{pash88} V.V. Pashkevich, Nucl. Phys. A \textbf{477}, 1 (1988).
\bibitem{rabotnov} A.S. Stavinsky, N.S. Rabotnov, and A.A. Seregin, Yad. Fiz. {\bf 7}, 1051 (1968)
\bibitem {brdapa} M. Brack, J. Damgaard, A.S. Jensen, H.C. Pauli, V.M. Strutinsky and C.Y. Wong, Rev. Mod. Phys. {\bf 44}, 320 (1972).
\bibitem{KarlHeinz} K.-H. Schmidt, J. Benlliure, A.R. Junghans, Nucl. Phys. A \textbf{693}, 169 (2001).
\bibitem{SS-Nature} G. Scamps and C. Simenel, Nature (London) \textbf{564}, 382 (2018).
\bibitem{CSN} N. Carjan, A.J. Sierk, J.R. Nix, Nucl. Phys. A \textbf{452}, 381 (1986).
\bibitem{wada21} T. Wada, K. Okada, N. Carjan, and F.A. Ivanyuk, EPJ Web of Conferences {\bf 256}, 00018 (2021).
\bibitem{Pyatkov} Yu.V. Pyatkov, D.V. Kamanin, N. Carjan, K. Okada, Z.I. Goryainova, E.A. Kuznetsova, V.D. Malaza, A.O. Strekalovsky, O.V. Strekalovsky, S.M. Wyngaardt, V.E. Zhuchko, Journal of Physics: Conference Series 2586, 012038 (2023).
\bibitem{viola1985} V.E. Viola, K. Kwiatkowski, and M. Walker, Phys. Rev. C {\bf 31}, 1550 (1985).
\end{thebibliography}
\end{document}